\input{aipcheck}

\documentclass[
    ,final            
  ]
  {aipproc}
\newcommand{\be}{\begin{equation}}
\newcommand{\ee}{\end{equation}}
\newcommand{\bea}{\begin{eqnarray}}
\newcommand{\eea}{\end{eqnarray}}
\newcommand{\nnu}{\nonumber\\}
\newcommand{\oot}{\overline {126}}
\newcommand\sigb{\bar\sigma}
\newcommand\ytt{\tilde {y}}

\newcommand\mzt{{\tilde{m}}_0}
\newcommand\Azt{{\tilde{A}}_0}
\newcommand{\eq}[1]{\begin{equation}#1\end{equation}}

\newcommand{\subs}[1]{_\mathrm{#1}}

 \newcommand{\mpl}{M\subs{p}}
\layoutstyle{8x11single}

\begin{document}

\title{Susy Seesaw Inflation and  NMSO(10)GUT }


\author{Charanjit S. Aulakh}{
  address={Dept. of Physics, Panjab University, Chandigarh, India,
  160014}}

\begin{abstract}
We show that Supersymmetric  models  with   Type I seesaw neutrino
masses support slow roll inflection point inflation. The inflaton
is the D-flat direction     labelled by the chiral invariant  HLN
composed of the Higgs(H), slepton(L) and conjugate sneutrino(N)
superfields. The scale of inflation and fine tuning is set by the
conjugate neutrino Majorana mass $M_{\nu^c} \sim 10^6-10^{12}$
GeV.   The cubic term in the (quartic) inflaton potential  is
dominantly from superpotential (not soft Susy breaking) couplings.
The tuning conditions are thus  insensitive to  soft supersymmetry
breaking parameters and are generically much  less stringent than
for previous `A-term'  inflation scenarios controlled by mass
scales $\sim TeV$. WMAP limits on the ratio of tensor to scalar
perturbations limit the scale $M$ controlling inflection point
inflation: $M <7.9 \times 10^{13}$ GeV.  `Instant preheating' is
operative and dumps the inflaton energy into MSSM modes giving a
high reheat temperature : $T_{rh} \approx
M_{\nu^c}^{\frac{3}{4}}\, 10^{6}$ GeV $\sim 10^{11}- 10^{15} $
GeV.    A large gravitino mass  $> 50 $ TeV is therefore required
to avoid over closure by reheat produced gravitinos. `Instant
preheating' and NLH inflaton facilitate production of right handed
neutrinos during inflaton decay and thus non-thermal leptogenesis
in addition to thermal leptogenesis. We show that the embedding in
the fully realistic New Minimal Supersymmetric SO(10) GUT requires
use of the heaviest righthanded neutrino mass as the controlling
scale but the possibility of a measurable tensor scalar
perturbation ratio seems marginal. We examine the parametric
difficulties remaining.
\end{abstract}

\maketitle


\section{Introduction}
Primordial inflation is now generally accepted as the only viable
mechanism for setting the initial conditions for   Big bang
Cosmogony in a way compatible with the extreme CMB homogeneity
observed by   increasingly accurate satellite maps~\cite{COBE} of
the Microwave sky. The parameters of inflation
($P_R,n_s,{\cal{D}}_k(n_s)$) measured so far can be accounted for
by most of the many slow-roll inflation models proposed. With few
exceptions these models use inflaton(s) that have no role to play
outside of inflation since they have no connection with the known
fields of Particle Physics. Such a connection is however necessary
since the post-inflationary epoch must include   reheating phases
where the inflaton energy is converted into the matter and
radiation observed today. Models driven   by an inflaton composed
of SM\cite{masina},MSSM~\cite{MSSMflat} or GUT\cite{guth} fields
thus carry an obvious appeal.   Models of the second type type are
typically based on slow roll inflation along  ``D-flat
directions''  in the MSSM field space and these are conveniently
labelled by  holomorphic gauge invariants formed from
 chiral superfields.
Such models ( also called ``A-Term Inflation'' models
\cite{akm,hotchmaz,lythdimo}) typically  require extreme  fine
tuning between the soft terms to ensure an inflection or saddle
point of the field potential where the vacuum energy density
drives a burst of inflation but nevertheless allows ``graceful
exit'' due to the absence of a local minimum and the  associated
potential barrier which would prevent exit.
  Thus while they answer some of the relevant issues they have much scope for improvement.

In \cite{akm,hotchmaz} an A-term inflation model    was based on
small  neutrino yukawa couplings needed for realistic Dirac light
neutrino masses. The inflaton field was a gauge invariant $D$-flat
direction, $N H L$, where $N$ is the right handed sneutrino, $H $
is the MSSM Higgs doublet which gives masses to the up-type
quarks, and $L$ is the slepton field. When coupled with   soft
trilinear and bilinear supersymmetry breaking terms of mass scale
$\sim 100 ~GeV$  to $10~  TeV $  the associated
{\it{renormalizable}} inflaton potential can   be fine tuned to
achieve inflection point inflation consistent with
 WMAP 7 year data\cite{akm,hotchmaz}.

 The Type I seesaw\cite{seesaw}) mechanism
offers a more attractive explanation for small neutrino(Majorana)
masses ($m_\nu\sim (m_{\nu}^D)^2/M_{\nu^c}$)
  based on large right handed neutrino  masses $M_{\nu^c}>>M_S$.
It is natural to ask if theories with supersymmetric Type I seesaw
masses also support  inflation. The popular
Leptogenesis\cite{leptogen} scenario as well as the realistic Susy
Minimal SO(10)GUT strongly hint  at right handed neutrino masses
in the range $10^{6} $ to $10^{12}$ GeV. So for $V_{B-L}\sim M_X>
10^{16}$ GeV the superpotential couplings $f_A, A=1,2,3$  which
generate   $M_{\nu^c_A}\sim f_A V_{B-L}$, are very small ($
f_{A}\sim 10^{-9}$ to $10^{-4})$ and can also give rise to a cubic
term in the quartic inflaton potential. Thus the required
ingredients for inflation are already present in    Supersymmetric
Type I seesaw models.

Issues regarding natural values for superpotential couplings come
into focus when viewed in the context of the so called Minimal
Left Right supersymmetric models\cite{MSLRMs} and their embedding
in GUT models\cite{rparso10,MSGUTs}. SUSY Left-Right Models are
advantaged due to their protection of R-parity as a gauged
discrete symmetry, which provides a stable lightest supersymmetric
particle (LSP) which has the properties required to serve as WIMP
dark matter. They simultaneously and naturally implement Seesaw
mechanisms for neutrino masses\cite{MSLRMs}. Moreover such models
have also been incorporated in the realistic and predictive New
Minimal Susy SO(10)  grand unified
theories(NMSGUT)\cite{NMSGUTs,nmsgut3} where all the hard
parameters of the MSSM are fitted in terms of fundamental
parameters of the GUT and soft SUSY breaking parameters (of the
Non-Universal Higgs masses (NUHM) type) defined at the Unification
scale $M_X\sim 10^{16}- 10^{18} $ GeV. Such GUTs have viable Bino
dark matter candidates and make distinctive predictions for the
type of SUSY spectra observable at the LHC. In 2008, well before
 the discovery of Higgs mass of around 125 GeV
in 2011-2012  and the consequent realization that a general
framework such as the  phenomenological MSSM (pMSSM) requires that
the soft trilinear couplings $A_{t,b}$ be \emph{large}, we
concluded\cite{NMSGUTs} that \emph{the NMSGUT would  be falsified
} by its failure  to fit the down type quark masses \emph{unless}
$A_0,\mu$ were in the tens/hundreds  of TeV : leading to a mostly
decoupled mini-split supersymmetry type superspectrum with only
the LSP, gauginos and possibly a light slepton in the sub-TeV
range. The experimental data has now forced this realization on
practitioners of MSSM parametrology\cite{djouadi}. In the NMSGUT
it was a prediction.

Taken together with the possibility of small values for the light
generation Yukawa Dirac  couplings it is possible to implement
viable inflection point inflation by suitable tuning
 at the supersymmetric level itself. This is technically more appealing
than a tuning applied to soft susy parameters which, being
unprotected by SUSY, are unstable. We examine the reheating
dynamics briefly. We then derive derive the embedding of SSI in
the NMSO(10)GUT and the necessary  tuning conditions  and show how
to satisfy them explicitly.

\section{Generic properties of  Renormalizable Inflection point inflation}

In this section we outline the essential features of   inflection
point inflation deriving from a  \emph{quartic}   potential of a
single complex scalar field $\varphi$.    Since the  angular
degree of freedom   has positive curvature
 and cannot support inflation  one may assumed it fixed at its minimum by relaxation and focus on the
 remaining real field $\phi$    whose potential is generically
$ V= {\frac{h^2}{12}} \phi^4  -  {\frac{A h}{6\sqrt{3}}} \phi^3
  + {\frac{M^2}{ 2}}\phi^2 $.

  The fine-tuning $A=4M$ implies that one has a saddle point
  ($V'(\phi_0)=V''(\phi_0)=0$) at the field value $\phi_0= \frac{\sqrt{3} M}{h}$.
   So one  defines  fine-tuning parameter $\Delta$ through  $A=4 M{\sqrt{
1-\Delta }}$ ($\Delta=\beta^2/4$ in the notation of
\cite{hotchmaz}). The inflection point($V''(\phi_0)=0$) is at $
\phi_0 =\frac{\sqrt{3} M}{h}(1-\Delta +O(\Delta^2))$.  For small
$\Delta$  \bea V(\phi_0)&=&V_0=\frac{M^4}{4 h^2} (1 + 4
\Delta)\qquad;\qquad V'(\phi_0)=\alpha=\frac{\sqrt{3} M^3 \Delta}{
h} \qquad;\qquad V'''(\phi_0) = \gamma=\frac{2 M h}{\sqrt{3}}(1-2
\Delta)\label{leadingV}\eea  If   $h $ is tiny $V_0\ >> M^4$ and
$\phi_0>>M/h$;  $\gamma$ is
 small with  $h$, while $\alpha$ is small
by   tuning. Large vacuum energy and flatness around $\phi_0$ and
starting $\phi$ near  $\phi_0$ with $\dot \phi<< \phi_0^2 $ imply
the universe   executes slow roll inflation as  $\phi$ rolls
through an interval of width $\Delta \phi \sim V_0/\gamma M_p^2$
below $\phi_0$.
 The standard slow roll parameters  are  defined as($ M_{p }= 2.43 \times 10^{18} \,GeV$)  \bea
 \eta(\phi) = \frac{M_p^2 V''}{V} \simeq {\frac{M_p^2}{V_0}}\gamma
 (\phi-\phi_0)~~;~~
  \epsilon(\phi) = \frac{M_p^2}{2} (\frac{V'}{V})^2 \simeq(\alpha
+\frac{\gamma}{2}(\phi-\phi_0)^2)^2({\frac{ M_p^2}{2V_0^2}}) ~~;~
  \xi = \frac{M_p^4V' V'''}{V^2} \simeq \frac{M_p^4 \alpha \gamma}{V_0^2}
  \label{slrlprm}\eea

The observed CMB is a combined spectrum of modes which exited
 the horizon during inflation. We
approximate  it as  a single spectrum from a
representative(``pivot'')  mode that exits the co-moving  horizon
when $ \phi=\phi_{CMB}$. This is the field value near $\phi_0$
where the inflation giving rise to observable effects today kicks
in (when $N_{CMB}$ e-folds of inflation are remaining).  The power
spectrum and spectral index we see today are then
$P_R(\phi(N\subs{CMB}))$ and $n_s(\phi(N\subs{CMB}))$
respectively.
  The small first and third Taylor coefficients
$\alpha,\gamma$  determine\cite{lythdimo,lythstew,liddleleach} the
measured parameters of inflation ($P_R,n_s$) once the field values
($\phi_{CMB},\phi_{end}$) at the time of horizon  entry of the
``pivot'' momentum scale ($k_{pivot}=0.002$ Mpc${}^{-1}$) and at
termination of the slow roll are
fixed\cite{lythdimo,liddleleach}(on the basis of an overall
cosmogonic scenario and the consistency of the slow roll
approximation ($\eta(\phi_{end})\approx 1$) respectively). The
observable  number of e-folds   $N_{CMB}=N(\phi_{CMB})$ is the
number  of e-folds of inflation left to occur after  $\phi$
crosses $\phi_{CMB}$ (the field value when    the representative
primordial  fluctuation length  scale ($l_{pivot}=k_{pivot}^{-1}$)
becomes larger than the comoving horizon (  $1/a_k H_k$)).
Plausible inflationary cosmogonies require $ 40<N_{CMB} < 60$ and
this severely restricts the inflation exponents.

The slow roll inflation formula for the power spectrum of the mode
that is leaving the horizon when the inflaton rolls to $\phi$ and
the corresponding spectral index and it's variation with momentum
is(\cite{lythstew}) \be P_R(\phi)=\frac{ V_0}{24 \pi^2
\mpl^4\epsilon(\phi)}~~;~
  n_s(\phi) \equiv
 1+2\eta(\phi)-6\epsilon(\phi)~~;~
   {\cal D}_k(n_s) = \frac{kdn_s(\phi)}{dk} =-16\epsilon\eta + 24
\epsilon^2 + 2 \xi^2\ee    In practice $\epsilon,\xi$ are so small
in the narrow region near $\phi_0$ where slow-roll inflation
occurs that their contribution to $n_s$ is negligible.   ${\cal
D}_k(n_s)$ is negligible i.e. the spectral index is scale
invariant in the observed range, as is allowed by observation so
far.

The field value at the end of   inflation $\phi_{end}$ is defined
by  $ \eta(\phi_{end})\simeq 1$  which gives
$\phi_0-\phi_{end}=\frac{V_0}{\gamma M_p^2}$. In the slow roll
approximation $ \dot\phi=-V'(\phi)/3 H >>\ddot\phi /H$,
 where $H=\sqrt{V(\phi_0)/(3 M_p^2)}$ is the (constant) inflation
  rate during slow roll inflation. One has the $N-\phi$ link (which
  can be exactly inverted\cite{ssi},without assuming that $\phi_{end}<<\phi(N)$
\cite{lythdimo})):
   \be  N(\phi) =  -3 \int_{\phi}^{\phi_{end}}\frac{H^2}{V'(\phi)}
   d\phi
    =  \sqrt{\frac{2}{ \alpha\gamma}}\frac{V_0}{M_p^2}
    \big(\arctan\sqrt{\frac{\gamma}{2\alpha}}(\phi_0-\phi_{end})-
    \arctan\sqrt{\frac{\gamma}{2\alpha}}(\phi_0-\phi
    )\big)\label{Nphi}\ee

  $N_{pivot}$ is estimated  using the standard Big Bang thermal cosmogony.
giving \cite{liddleleach} $ N\subs{pivot} =
 65.5+ \ln\frac{\rho_{rh}^{\frac{1}{12}}V_0^{\frac{1}{6}}}{\mpl }
 $ where $\rho_{rh}$ is the energy density after reheating and $V_0$
the potential value during inflation. Due to rapid thermalization
in this model(see below) the two are equal and then since the
scale is set by $V^{1/4}\sim M/\sqrt{h}$ one finds $N_{pivot}=46~$
to $~55=51\pm 5$ to be a reasonable estimate.

To search for sets of   potential parameters
 $M,h,\Delta$ compatible with the observed  $P_R,n_S,N_{CMB} $  one uses the
        definitions
  \be  \epsilon_{CMB} = \frac {V_0}{24 \pi^2 M_p^4 P_R}\qquad;\qquad
  \eta_{CMB}  =  \frac{(n_s-1)}{2}\label{epsetacmb}\ee

and from these deduces
  $\alpha_{CMB},\phi_{CMB}$    using the
 eqns.(\ref{slrlprm}) \be \phi_{CMB} =  \phi_0 + {\frac{V_0 \eta_{CMB}}{ \gamma  M_p^2}} \qquad;\qquad
\alpha_{CMB} =  \sqrt{2  \epsilon_{CMB}} {\frac{V_0}{M_p}}
-{\frac{V_0^2\eta_{CMB}^2}{2 \gamma  M_p^4} } \ee

 The required fine-tuning $\Delta$  is then  \be \Delta= {\frac{h\alpha_{CMB}}{\sqrt{3}M^3}}=(\frac{M}{4 h M_p})^4
 (\frac{16 h^2 M_p}{3 \pi M {\sqrt{P_R}}}-(1-n_s)^2)\ee
 $\alpha_{CMB},\Delta$ should emerge real and positive
 and using $\{\alpha_{CMB},\phi_{CMB}\}$ in the formula for
$N_{CMB}$ one should obtain a sensible value  in the range
$N_{CMB}=51\pm 5$. Using eqns.(\ref{leadingV},\ref{epsetacmb}) in
eqn(\ref{Nphi})  we can solve accurately for the required relation
between $h,m,\Delta$ using an interpolating function \cite{ssi}.
The result is   that  $N_{CMB} \sim 50$, $Z_{0}\approx
\frac{1.2}{N_{CMB}}$ solves the exact equations  to a good
approximation and one obtains the generic constraints : \be
\frac{h^2}{M}  \approx  \frac{3 \pi}{M_{P}}
\frac{\sqrt{P_R}}{N_{CMB}^2} \approx \frac{2.75 \times
{10^{-22}}}{N_{CMB}^2} \approx 10^{-25}~GeV^{-1}\qquad;\qquad
\frac{\Delta}{M^2} \approx \frac{4.14 \times 10^{-34}}{N_{CMB}^2
P_R} \approx 10^{-28.2} GeV^{-2}\label{hsqbyM}\ee
 We then have viable inflation with inflaton energy and Hubble rate \be V_0
\sim \frac{M^4}{h^2}\sim (M)^3\times 10^{25} \, GeV \sim
10^{43}-10^{61} \, GeV^4 ~~~~~~;~~~~~~ H_0 \sim
\sqrt{\frac{V_0}{M_P^2}} \sim 10^{3 }-10^{12} \, GeV  \ee The
fine-tuning measure grows as $M^2$ so that $\beta=\sqrt{\Delta}$
can be as large as $10^{-2}$ for $M\sim 10^{12}$ GeV. In our
scenario due to the large value of the inflaton mass parameter
$M\sim 10^6 ~ $to $~ 10^{13}$ GeV compared to $M\sim TeV$ in  the
case of  MSSM inflation\cite{MSSMflat} or Dirac neutrino
inflation\cite{akm,lythdimo}  the   fine-tuning
  required is quite mild and removes much of the motivation
   for complicated just so hybrid inflation scenarios.

We can also estimate the ratio $r$ of  power in Tensor and Scalar
CMB fluctuations using $r=2 V_0/(3 \pi^2 P_R M_P^4)$. On    using
  eqn.(\ref{hsqbyM}) \be r=\big( \frac{M}{7.95\times 10^{13} GeV}\big
  )^3\ee
  This makes the observation of tensor perturbations in such a
  scenario hard unless $M$ is near its upper limit.

\section{  Supersymmetric   seesaw Inflaton model }

In this section we introduce a toy one generation  Supersymmetric
seesaw inflation scenario   model with gauge group $ SU(3)\times
SU(2)\times U(1)_R\times U(1)_{B-L} $ that captures the essential
features of our scenario.  The essential fields beyond the MSSM
are a right handed Neutrino chiral multiplet   $N [1,1,-1/2,1]$
and a field ${ S}[1,1,1,-2]$ whose vev generates the large
Majorana masses $M_\nu$ ($10^6-10^{14}$ GeV) for the conjugate
neutrinos $\nu^c_A \equiv N_A$ via a renormalizable superpotential
coupling $3 \sqrt{2} f_{AB} S \nu^c_A \nu^c_B   $.
  Additional superheavy fields $\Omega_i$   serve to fix
 the vev of $S$   as in Minimal Supersymmetric
 Left Right Models (MSLRMs)\cite{MSLRMs} and in GUTs that embed them
\cite{rparso10,MSGUTs,NMSGUTs}. Neutrino  Dirac Yukawa coupling is
present in the superpotential :  $W=y_{\nu}N L  H +...$ where
$L[1,2,0,-1],H[1,2,1/2,0]$ are the Lepton doublet and up type
Higgs respectively. The relevant D-flat direction extends out of
the minimum of the supersymmetric potential corresponding to the
breaking of the gauge group down to the MSSM symmetry  \bea
SU(3)\times SU(2)\times U(1)_R\times U(1)_{B-L} \rightarrow
SU(3)\times SU(2)\times U(1)_Y\eea This leads to a Type I seesaw
plus MSSM (SIMSSM) effective theory. After symmetry breaking the
MSSM hypercharge  $Y= 2  T_{3R}  + (B-L)$ where $T_{3R}$ is the
$U(1)_R$ generator. Unlike the case of the Dirac neutrino masses
scenario \cite{akm} $B-L$ is \emph{not} a gauge symmetry down to
low energies. This can have important consequences for
nucleosynthesis and matter domination since the heavy right handed
neutrinos must find a non-gauge channel to decay through. The
flat-direction associated with the gauge invariant $NLH$ is then
specified as
\begin{equation}
\tilde{N}= \tilde{\nu}=h_0=\frac{\varphi}{\sqrt{3}}=\phi
e^{i\theta};~~~\phi\geq 0,~~ \theta\in[0,2 \pi)\, \label{inflaton}
\end{equation}
The additional fields $\Omega_i$ are   coupled to $S$ so that
extremization of the SUSY potential using
$F_{\Omega_i}=0,~~D_\alpha|_{\phi=0}=0 $ fixes the vev of S:
$<S>=\bar\sigma/\sqrt{2}$ without constraining the inflaton field
$\varphi$. This is as in the Minimal Susy LR models\cite{MSLRMs}
and renormalizable  Susy SO(10) GUTs \cite{MSGUTs,NMSGUTs} which
are our inspiration and target.

At scales $\phi\sim \bar\sigma>>M_{S}$
 where SUSY is exact the relevant superpotential is given by:
\begin{equation}
    W=    3\sqrt{3} y N   \nu h+ 3 f{\sqrt{2}} S NN +...= y\varphi^3 +f{\sqrt{2}}
     S  \varphi^2 +...
\label{WSIMSSM}\end{equation}
where $ h,f,\bar\sigma$ can be taken real without loss of
generality.The equations of motion of the unperturbed vacuum imply
$<F_S>=0,  <S>= \bar\sigma/\sqrt{2}$. The right handed neutrino
Majorana mass will be $M_{\nu^c}=6f\sigb $.

This superpotential leads to an inflaton   potential
\begin{eqnarray}
    V_{susy}&=&|3 y \varphi^2 + 2 f\bar\sigma\varphi|^2 +   2 |f
    \varphi^2|^2\nonumber\\
    &=& f^2     \left[ (2+9{\tilde{y}}^2){\phi}^4 +12
    {\tilde{y}}{\phi}^3 \sigb\cos\theta + 4 \sigb^2{\phi}^2\right]
\end{eqnarray}
Here $\tilde{y} =y/f$ and  $f\sigb$ sets the mass scale.
Minimizing with respect to $\theta$ gives $\theta=\pi$ so we can
focus on just the real part of $\varphi$ and set
 $\varphi=-\phi$ with $\phi$ real and positive near the inflection
 point but free to fall into the well around $\phi=0$  and
 oscillate around that value.
  In addition one also expects a contribution to the potential
  from the $\mu $ term for the Higgs doublets together with
SUSY breaking quadratic and cubic soft terms, which we assume to
be of the type generated by supergravity, but with non universal
Higgs masses,  i.e of the form:
\begin{eqnarray}
    V_{soft} &=& \big [A_0 (y \varphi^3  + f \sqrt{2} S\varphi^2)+ h.c\big ] +m_{\tilde f}^2
    \sum_{\tilde f} | {\tilde f}|^2 + m_{ H}^2 |  H|^2  + m_{\bar H}^2 |\bar H|^2  \nnu
    &=& f^2    \left[\ytt\Azt\phi^3 \sigb \cos{3\theta} +  \Azt \sigb^2{ \phi}^2
    \cos{2\theta}+\mzt^2 \sigb^2 { \phi}^2 \right]
    \end{eqnarray}
here $\mzt=m_0/{f\sigb}, \Azt= 2 A_0/{f\sigb} $. The soft mass
 $m_0$ receives contributions from the sfermion and Higgs soft
masses as well as the $\mu$ term : $ m_0^2= ({2 m_{\tilde f}^2 +
\overline{m}_H^2})/{3}$.  Here $m_{\tilde f,H}$ are the sfermion
and up type Higgs soft effective masses at the unification scale
($\overline{m}_H^2=m_H^2+|\mu|^2$). Since these masses and $A_0$
should be in the range $10^2-10^5$ GeV while the righthanded
neutrino masses lie in the range $10^{6}-10^{12}$ GeV, it is clear
that $\tilde m_0,\tilde A_0 $ are   small parameters and even for
the large values of $m_0, A_0\sim 10^5$ GeV found in the NMSGUT
$\mzt,\Azt <<1 $.  Thus these terms cannot significantly change
$\theta=\pi$ assumed earlier. The total inflaton potential is then
\eq{V_{tot}=f^2\left((2+9\tilde y^2) \phi^4-(\tilde A_0+12)\tilde
y\bar\sigma\phi^3 +  (\tilde A_0+\tilde
m_0^2+4)\bar\sigma^2\phi^2\right).\label{eq-inf-Vtot}} Thus we
have  a generic quartic inflaton potential of the same type as in
Section $\bf{2}$ but  the parameter values  in the case of Type I
seesaw are quite different from the light Dirac neutrino case.  We
  have the  identification
of parameters \bea h&=&f\sqrt{12(2 + 9 \ytt^2)}\nonumber\\
A&=&\frac{3 f (\Azt +12)\ytt \sigb}{\sqrt{(2 + 9
\ytt^2)}}\nonumber\\
M^2&=& 2 f^2 \sigb^2(4+\Azt +\mzt^2)\nonumber \\
\Delta&=& (1-\frac{A^2}{16 M^2})\nonumber\\
&=&\left(1-\frac{9\tilde y^2(\tilde A_0+12)^2}{32(2+9\tilde
y^2)(\tilde A_0+\tilde m_0^2+4)}\right)\label{paramident}\eea For
seesaw models the natural magnitude for the neutrino Dirac mass
is, $m_{\nu}^D >1 MeV $ (i.e $\, |y_\nu^D| > 10^{-5}$ and then the
limit $m_{\nu}<<0.01 eV$ for the lightest neutrino (assuming
direct hierarchy) implies $M_{\nu^c} > 10^6$ GeV). Since the
preferred values for the Susy breaking scale are smaller than 100
TeV (at most) it follows that the maximum value of $|\Azt|,|\mzt|
\sim 0.1 $ and they could be much smaller for more typical larger
values of the conjugate neutrino masses $M_{\nu^c} \sim 10^8 $ to
$ 10^{12}$ GeV.  It is then clear from the corresponding range
$\Delta\sim 10^{-12} $ to $10^{-4}$ that the coupling ratio
$\ytt=y/f$ becomes ever closer to exactly $\ytt =4/3$ as M
increases and even for $M\sim 10^6$ GeV differs from $1.333 $ only
at the second decimal place. Thus to a good approximation $
h=6{\sqrt{6}} f $. Then it follows  that \be  f
   \simeq
10^{-26.83 \pm 0.17}(\frac{\overline{\sigma}}{GeV})~~;~
 M \simeq 10^{-25.38 \pm
 0.17}(\frac{\overline{\sigma}}{GeV})^2~~;~
 \Delta  \simeq   10^{-78.93 \pm 0.47}
 (\frac{\overline{\sigma}}{GeV})^4\ee
 The range $M\sim
10^{6.6}$ to $10^{10.6}$ GeV  corresponds nicely to  $10^{16}
~GeV< \sigb < 10^{18} ~GeV $:  as is natural in single scale Susy
SO(10) GUTs\cite{rparso10,MSGUTs,NMSGUTs,nmsgut3}.  $f$ increases
with $\sigb$  with values below   $10^{-11}$ achievable in the
NMSGUT only with difficulty.  In MSLRMs, since there are no GUT
constraints on $\sigb$, one may  assume somewhat wider ranges for
these parameters.

 In all relevant cases $\Delta < 10^{ -4}$ is required. Thus the above equations imply
that ${\tilde y}^2$ must be close to the value \be {{\tilde
y}_0}^2 = \frac{64}{9}  {\frac {4+\Azt +\mzt^2}{16-8\Azt-32
\mzt^2+\Azt^2}}\ee
  Here $\Azt,\mzt \sim O(M_S/M_{\nu^c})
<<1$,  hence $\tilde y_0$ is rather close to $4/3$ and the
equality is very close for larger $M\sim f\sigb$ since then
$\Azt,\mzt$ are tiny.
 The measure of severity of
fine tuning $\beta=\sqrt{\Delta} \sim 10^{-2} - 10^{-6}$ compares
quite  favourably with the case of the MSSM or Dirac neutrino
inflaton since there $\beta\sim 10^{-12} $ to $10^{-10}$ due to
the low values of the inflaton mass in those cases. The dominant
component of the fine tuning in the present case  is a fine-tuning
of superpotential parameters, which is radiatively stable due to
non renormalization theorems. Specially for large $\sigb > 10^{16}
~GeV$ the Type I Susy seesaw can provide a rather attractive
inflationary seesaw with a natural explanation for neutrino masses
and   weaker tuning demands on the radiatively unstable Susy
breaking parameters than the extreme and unstable fine-tunings
demanded by typical inflection point scenarios and in particular
the Dirac neutrino model \cite{akm}. Moreover, unlike the chaotic
sneutrino inflaton scenario\cite{murayana,elliyana}, no
trans-Planckian vevs are invoked.

\subsection{Reheating and Leptogenesis}

The post inflationary dynamics of our model bears an intimate
relation to previous studies of models with`instant preheating'
mechnism\cite{feldkoflinde} and specially the   MSSM flat
direction inflection point inflation   model\cite{alfermaz} and
preheating model\cite{ahnkolb} with strong coupling to the MSSM
Higgs.   Supersymmetric seesaw inflation offers an attractive
synthesis precisely fulfilling the need expressed in
\cite{ahnkolb} :

 "\emph{There
have been many models of leptogenesis. A hallmark of our model is
the economy of fields. The only undiscovered fields are the
inflaton, $\phi$, the standard model Higgs, h, and the
right-handed neutrino, N. There are very good reasons for
suspecting that all exist! The only unfamiliar aspect of our model
is the strong coupling of the inflaton field to the Higgs field.
While there is no reason to preclude such a coupling, it would be
very interesting to find particle-physics models with a motivation
for the coupling. }"

   Due to the gauge(H,L)  and third generation yukawa($H$)
    coupled components of the inflaton
the inflaton energy will decay very rapidly (with decay time
$\tau_{dec} << H_{infl}^{-1} \sim (h M_p)/M^2$)  by the `instant
preheating'' mechanism\cite{feldkoflinde,ahnkolb,alfermaz}.  Thus
 the  reheating temperature $ T_{rh}\sim T_{max} \sim V_0^{1/4}  \sim
M/h^{1/2}\sim 10^{11}-10^{15}\, GeV$. The parametric dependence is
identical to  that found in \cite{alfermaz}. The difference in
scales arises only because the inflaton mass $M\sim 10^6-10^{12}$
GeV  in our model is much larger than the inflaton mass parameter
$m_\phi\sim 0.1-10$ TeV  in \cite{alfermaz}  coming from soft
Supersymmetry breaking.

 In  preheating   (``$\chi $ type")  degrees of freedom, with masses($m_\chi\sim
g \phi(t)$) and decay rates ($\Gamma \sim g^3 \phi(t)$)
proportional to  $\phi(t)$,  are produced non-perturbatively every
time the inflaton field crosses zero. This happens because the
$\chi$ modes are ultra-light for a sufficiently large time
interval around the zero crossing time during which adiabaticity
is violated ( $ {\dot\omega}_k  > \omega_k^2$ : where $\omega_k$
is the oscillation frequency at wave number $ k$). Here $\chi$
modes are the components of the $H,L,u^c_L,u_L$ chiral superfields
and the $W_\pm,B$ gauge superfields. They can be identified  as
the fields which become massive given  background values of the
three components of the inflaton ($\tilde \nu,\tilde
\nu^c_L,h^0$). Then with the usual superpotential
 \be W=y^u Q_L H u^c_L  + y^d Q_L {\overline H}
  d^c_L + y^\nu L H N  + y^l L{\overline H}e^c_L +...\ee
  $y^u$ leads to massive $u_L,u^c_L$;  $y^\nu$ leads to
massive $e_L$(one combination of the three $e_L$ ), $h^0,h^+
,\nu_L,\nu^c_L$; $y^l$ leads to massive ${\bar h}^-,e^c_L$(one
combination). Since $<H,N,L>$ preserve $U(1)_{em}$, the gauge
couplings give masses to Z (which forms a Dirac supermultiplet
with $(\nu-{\tilde h}_0)/\sqrt{2}$) and $W_{\pm} $ (form a pair of
Dirac supermultiplets with $l^{-},{h}_+$).   The inflaton vev
leaves the down quark and gluon/gluino fields and $\bar h_0$, and
some combinations of the $l^-_L,l^c_L$ fields with light (MSSM
type) masses. These light ($\psi$-type) fields will form the first
step in the decays of the $\chi$ field.
   As $<\phi>$ again increases after crossing zero  the
$\chi$ modes become   heavy and unstable and as a result decay
rapidly(within a time $\tau_{dec}\sim \frac{h}{M g^3}<<
m_{\phi}^{-1}$)  to the light (mostly coloured) MSSM modes  to
which they are coupled strongly coupled. A
 fraction $\sim 10^{-1}$ of the inflaton condensate energy  passes into
 the light MSSM modes with every crossing resulting in complete
 transfer within $\sim 10^2$ oscillation times. $\tau_{osc}\sim m_{\phi}^{-1}<<
 H_{infln}^{-1} \sim (h M_p)\tau_{osc}/M \sim (1  \, - \, 150)
 \tau_{osc}$.  Once the energy is in the light($\psi$) modes  MSSM
 interactions rapidly complete thermalization.
Rapid decay of the inflaton oscillation amplitude   leaves the
light modes to thermalize the energy dumped by the inflaton into a
radiation bath of all modes: which are no longer ever heavy
because the inflaton has decayed.
 The reheating temperature is \bea T_{rh} \sim ({\frac {30} {\pi^2 g_*}})^{1/4} V_0^{1/4}
 \sim T_{max} \sim 10^{11} - 10^{15} \, GeV \eea
where $g_*=228.75$ is the effective number of MSSM degrees of
freedom. This  reheating temperature is well above that required
to produce relativistic  populations of gravitinos : which are
unacceptable if their lifetimes are  larger than the
nucleosynthesis  time $\tau_{N} \sim 1 \, sec $ since their decay
after nucleosynthesis would destroy the created nucleons. The
straightforward and generic resolution of this gravitino problem
is if the graviton masses are sufficiently large so that the
gravitinos decay before nucleosynthesis\cite{moroi}  : $
\tau_{grav} \sim 10^5 \, sec ({\frac{1 \, TeV }{m_{3/2}}})^3 <<
\tau_N \sim 1 ~ sec $.  Thus  Supersymmetric seesaw Inflation also
indicates that the scale of supersymmetry breaking -as indicated
by the gravitino mass- should  be  above $ 50 $ TeV;  as is also
found by fitting of fermion data in the NMSGUT\cite{NMSGUTs}.
Large reheat temperatures also ensure  abundant thermal production
of righthanded neutrinos after inflation. Their CP violating
decays into leptons can drive thermal  lepto-genesis
\cite{leptogen} for generating the observed baryon to entropy
density  $n_B/s \sim 10^{-10}$ .  \emph{Non-thermal} leptogenesis
is also possible \cite{ahnkolb} since   the Higgs field H is
itself a $\chi$ type field and  coupled to the righthanded
neutrinos. During inflaton  oscillations  the Higgs mass $m_h\sim
g_2 \phi$ fluctuates    below and above      $M_{\nu^c}\sim f
 {\bar\sigma} $.  CP violating Higgs-righthanded Neutrino
inter-conversion\cite{ahnkolb} leads to (non-thermal) Leptogenesis
which will add to the thermal leptogenesis. The complication in
the present case that the $L, H$ and $N$ components of the
inflaton  have   different decay rates implies a proper analysis
must track the separate evolution of all three fields making up
the inflaton using the  equation of motion and Boltzmann equation
for the relevant degrees of freedom.  This requires a separate
numerical study to expose  the  interplay of the couplings
$f_A,y_{AB},g_2$. The study of this evolution and the operation of
Leptogenesis in these models is now in progress.

\section{ Inflation and neutrino masses in the NMSGUT}

Finally we    embed   SSI  in the New Minimal SO(10) GUT
(NMSO(10)GUT or NMSGUT). The NMSGUT is a realistic Susy SO(10)
model\cite{NMSGUTs,nmsgut3,blmdm} that successfully fits the known
fermion mass-mixing data in terms of GUT parameters and provides
structural reasons for suppression of the dangerous operator
dimension $d=4,5$ Baryon violation typical in Susy
GUTs\cite{nmsgut3}. It furthermore makes distinctive predictions
of  a mini-split supersymmetry spectrum made viable by  large $A,
\mu$-terms and with a characteristic \emph{normal} s-hierarchy.
Neutrino flavour plays a key role in enabling NMSGUT inflation :
the   inflaton is composed of third generation conjugate
sneutrino, first generation left slepton (sneutrino) and the
$T_{3R}=1/2$ Higgs.

The NMSGUT  Higgs field vevs
$\{{\mathbf{210}}(\omega,p,a),\mathbf{126}(\sigma)\}\equiv \Omega,
\mathbf{\oot}(S=\bar\sigma)$  break $SO(10)\rightarrow G_{MSSM}$
while preserving Supersymmetry  at $M_X$. An explicit Susy
preserving solution of symmetry breaking in terms of a  cubic
equation for a complex variable $x$ and depending on a single
parameter ratio $\xi$ was found by us\cite{MSGUTs}.  The mass
spectra implied\cite{MSGUTs,NMSGUTs} by this analytic solution
for the the MSGUT vacuum   are the basis of our detailed
Renormalization Group and threshold effect
analysis\cite{MSGUTs,NMSGUTs,nmsgut3}. Inclusion of threshold
corrections   raises the  unification scale close to the Planck
scale and can lower the gauge coupling at unification. We shall
use the notation and results  of \cite{ag1,ag2,NMSGUTs,nmsgut3}.

   To embed SSI  corresponding to a  $NLH$ type  flat direction  we
   show there is  a corresponding flat direction of the full GUT potential
 which  rolls out of the    MSGUT minimum (that has the  SIMSSM as its effective theory).
  The relevant fields are the GUT scale  vev fields $\Omega\equiv \{\omega,p,a,w,\sigma\},S=\bar\sigma$  and
 the (6) possible components $ h_i,{\bar{h}_i};i=1...6$ of the light MSSM Higgs doublet
 pair $H,{\overline H} $ together with  the  chiral lepton fields $L_A,\nu_A^c,
 A=1,2,3$.  The relevant superpotential is
 then\cite{MSGUTs,NMSGUTs}
\be  W =  2\sqrt{2}(h_{AB}h_1-2\sqrt{3}f_{AB}h_2-g_{AB}(h_5+i
\sqrt{3}h_6))+\bar{h}^T {\cal{H}}(<\Omega>)h  +4\sqrt{2}
f_{AB}\bar\sigma  \bar \nu_{A} \bar\nu_B + W_\Omega(\Omega) \ee
where
  \be
 W_\Omega(\Omega,{\bar\sigma}) =  m(p^2+3a^2+6\omega^2)+2 \lambda(a^3+3p\omega^2) +(M+\eta(p+3a-6 \omega))\sigma \bar\sigma
 \ee
 and \be \frac{\partial W_{\Omega}}{\partial \Omega }|_{h,\bar \nu,L = 0} =
 \frac{\partial W_{\Omega}}{\partial \bar\sigma  }|_{h,\bar \nu,L = 0}=0
 ~~~~~~~~~~~~~~
  D_{\alpha}(\Omega)|_{h,\bar \nu,L = 0}=0\label{omegavac}\ee
 here  $h_{AB},g_{AB},f_{AB}$ are the yukawa coupling matrices of
the three matter 16-plets  to the $\mathbf{10,120,\oot}$ Higgs
multiplets respectively. ${\cal{H}}$ is the Higgs doublet mass
matrix\cite{ag2,MSGUTs,NMSGUTs}. Equation (\ref{omegavac}) defines
the MSGUT vacuum\cite{MSGUTs}.

Of the 5  diagonal D-terms of SO(10)  only those corresponding to
the generators $T_{3L},\break T_{3R},B-L $ are relevant for  vevs
  $\Omega,\bar\sigma$ and out of equilibrium inflaton mode composed of  $\nu,\nu^c,h_0$. The vevs $\Omega,\bar\sigma$ do not
contribute to these D terms or cancel  so
 \bea
 D_{3L} &=&\frac{g_{u}}{2}(-\sum_{i=1}^6|h_{i0}|^2+\sum_{A}|\tilde
 \nu_A|^2)\nnu
D_{3R}&=&\frac{g_{u}}{2}(\sum_{i=1}^6|h_{i0}|^2-2|h_{40}|^2-\sum_{A}|\tilde
{\bar\nu}_{A}|^2)\nnu
D_{B-L}&=&\sqrt{\frac{3}{8}}g_{u}(\sum_{A}(|\tilde
{\bar\nu}_{A}|-|\tilde \nu_A|^2)+2|h_{40}|^2)\eea where  only
$h_{4\alpha}=\Phi^{44}_{{\dot{2}}\alpha}$ has $B-L=+2,T_{3R}=-1/2
$ and thus $Y=1 $ while all others have $T_{3R}=1/2$ and $B-L=0$.
The D-flatness conditions are
  \be \sum_A|\tilde
\nu_A|^2=\sum_{i}|h_{i0}^2|=\sum_A |\tilde{\bar
\nu}_{A}|^2+2|h_{40}|^2 \ee In MSGUTs the MSSM Higgs doublet pair
is defined by fine tuning $Det({\cal{H}})\simeq 0$ so that its
lightest eigenvalue $\mu \sim M_W\sim 1 $ TeV specifies the $\mu$
term in the superpotential of the SIMSSM : $W=\mu
{\overline{H}}H+...$. The doublet pair $H,{\overline{H}}$ is a
linear combination\cite{MSGUTs,ag2,ag1} of the 6 doublet pairs of
the the NMSGUT :
 \bea  h_i=U_{ij} H_j   \qquad \qquad  \bar{h}_i={\overline{U}}_{ij}
 \overline{H}_j\eea
where $U,{\overline{U}}$   diagonalize the doublet mass matrix
${\cal{H}}$ : ${\overline{U}}^T{\cal{H}} U=
Diag\{\mu,M^H_2,....,M^H_6\}$ to positive masses. They are
calculated with $\mu=0=Det({\cal{H}})$. The so called Higgs
fractions : $\alpha_i=U_{i1},\bar\alpha_i={\overline{U}}_{i1}$ ,
  determine the grand unified
formulae\cite{MSGUTs,NMSGUTs} for the SIMSSM fermion yukawas. For
tree level yukawa couplings  replace  $ h_i,\bar{h}_i\rightarrow
\alpha_i H ,\bar{\alpha}_i {\bar{H}} $. For example   the neutrino
Dirac coupling is ($(\tilde h_{AB},\tilde g_{AB},\tilde
f_{AB})$=$2\sqrt{2}(h_{AB},g_{AB},f_{AB})$ ) \be
y^{\nu}_{AB}=\tilde h_{AB} \alpha_1-2\sqrt{3}\tilde
f_{AB}\alpha_2-\tilde g_{AB}(\alpha_5+i \sqrt{3}\alpha_6)\ee From
the $V=|F_{\bar h}|^2$ only the light Higgs doublet $H$  can
contribute. To get small yukawas   the involvement of the lightest
generation is unavoidable. Thus we take $\nu_{A}=\nu_1$. Taking
$\tilde{  \bar\nu }_{ A}=\tilde{\bar \nu}_{1}$  the tuning
constraint is  \emph{at best} of  form $|y_{11}|^2\sim 10
(|y_{21}|^2 +|y_{31}|^2)$ : this is impossible to  satisfy with
normal neutrino yukawa coupling  hierarchy.  Choosing
$\bar\nu_A=\nu^c_3 $ as the conjugate neutrino component of the
inflaton is more helpful in satisfying the fine tuning condition.
Thus our inflaton ansatz is
 \be \tilde{\nu}_1=
\frac{\phi}{\sqrt{3}} \qquad\qquad
 h_{i0}=\frac{\alpha_{i}\phi}{\sqrt{3}}  \qquad\qquad \tilde{\bar
\nu}_{3}=\frac{\phi}{\sqrt{3}}\sqrt{1-2|\alpha_4|^2}\ee Note how
the Higgs fraction $\alpha_4$   enters    as $\Gamma=1-
2|\alpha_4|^2$. It happens the solutions we have found earlier
\cite{NMSGUTs} can have $|\alpha_4|\sim 0.5$. It is not
inconceivable that $\Gamma\simeq 0$ is achievable without
destroying the realistic fermion fits to the fermion data.

We use   generic   Supergravity(SUGRY)-NUHM generated soft terms
in terms of a common trilinear parameter $A_0$  but
 different soft mass parameters $\tilde{m}_{\tilde
f}^2,\tilde{m}_{h_i}^2$ for the  16 plets  and the different
Higgs. Repeating the analysis of section 3 with the NMSGUT
superpotential and the new ansatz  we  obtain the parameter
identifications
  \bea
h&=&\frac{2}{\sqrt{3}}\big[(y^{\nu \dag} y^{\nu})_{11}+
\Gamma(|\tilde{h}_{31}|^2+4|\tilde
g_{31}|^2+(y^{\nu}y^{\nu \dag})_{33})+ 4|\tilde f_{33}|^2\Gamma^2)\big ]^{\frac{1}{2}}\nonumber\\
A&=&\frac{1}{h}(16|\tilde f_{33}|
|y^{\nu}_{31}||\bar\sigma|{\sqrt\Gamma} +
 4 |y^{\nu}_{31}|{\sqrt\Gamma}A_0 \cos (3
 \theta_{\bar\sigma}-2 \theta_{y^{\nu}_{31}}))\nonumber\\
M^2 &=&\frac{32}{3}|\tilde f_{33}|^2|\bar
\sigma|^2\Gamma+\frac{8}{3}A_0\tilde f_{33}|\bar \sigma| \Gamma
\cos (3\theta_{\bar\sigma}-2 \theta_{y^{\nu}_{31}})+2
{\widehat{m}}_0^2 \eea The fine tuning condition $A=4 M$  is now
 \be |y^{\nu}_{31}|^2 =  \frac{8
\Lambda_{n}}{9\Lambda_{d}-8 \Lambda_{n}(1+\Gamma)}\big
[|y^{\nu}_{11}|^2+|y^{\nu}_{21}|^2+\Gamma(|\tilde{h}_{31}|^2+4|\tilde
g_{31}|^2+|y^{\nu}_{32}|^2+|y^{\nu}_{33}|^2) + 4|\tilde
f_{33}|^2\Gamma^2\big ]\ee  Where \bea
\Lambda_{n}&=&1+\frac{A_0}{4M_3}\cos(3\theta_{\bar\sigma}-2
\theta_{y^{\nu}_{31}})+
\frac{3{{\widehat m}_0}^2}{16 M_3^2 \Gamma}\nonumber\\
\Lambda_{d}&=& (1+\frac{A_0}{4M_3} Cos(3\theta_{\bar\sigma}-2
\theta_{y^{\nu}_{31}}))^2\eea and $M_3=\tilde f_{33}|\bar \sigma|
$ $M_{\bar\nu}>>M_S$ imples  $\Lambda_{n,d}$ are both very close
to unity.  So as before the fine tuning condition is essentially
between hard parameters as in GUTs and in sharp contrast to MSSM
inflaton models\cite{MSSMflat}: \be |y^{\nu}_{31}|^2 = \frac{8
}{1-8\Gamma}(\Gamma(|\tilde{h}_{31}|^2+4|\tilde
g_{31}|^2+|y^{\nu}_{32}|^2+|y^{\nu}_{33}|^2) +
|y^{\nu}_{11}|^2+|y^{\nu}_{21}|^2+4|\tilde f_{33}|^2\Gamma^2)\ee
In NMSGUT fits the  strong hierarchy
$|y_{33}|>>|y_{32}|>>|y_{31}|>>|y_{21}|>|y_{11}|$ holds . So  one
must tune \be \Gamma \approx 0 \,\,\,\ {~{i.e}}\qquad\quad
 |\alpha_4|\approx {\frac{1}{\sqrt{2}}} \ee to a good accuracy. The
MSSM doublet H is almost exactly 50\% derived from the doublet in
the 210 plet ! The yukawa  tuning condition is only
\bea|y^{\nu}_{31}|^2&=& 8 (|y^{\nu}_{11}|^2+|y^{\nu}_{21}|^2 )\eea
which is easy to enforce in the NMSGUT.

There is an additional demand coming from eqn(\ref{hsqbyM}) :
$h^2/M_3 \sim ({y^\nu}^\dag y^\nu)_{11}/M_3 \sim 10^{-25} $ which
is, at first glance,   much harder to enforce.   However
\cite{nmsgut3}   large wave function corrections\cite{wright} to
the GUT($Y_f^{tree}$)-MSSM($Y_f$) yukawa coupling relation due to
the circulation of heavy fields within loops on the lines entering
the yukawa vertex imply  :
 \bea Y_f=(1 +{\Delta}_{\bar f}^T)\cdot (Y_f)_{tree}\cdot(1+  \Delta_f)
   (1+ \Delta_{H^{\pm}} ) \eea
 Due to the large number of
heavy fields the dressing of the Higgs fields can be rather large
($\sim 10^2 $). We  earlier  calculated\cite{nmsgut3} the
dressing for the \textbf{10}-plet component of the MSSM Higgs.
  Above  we showed  that a completely
independent line of argument \emph{requires} that the doublet $H$
be 50\% derived from the \textbf{210}-plet. Thus the lengthy
  calculation of the wave function corrections for each of the
 six GUT doublets  contributing   to the MSSM doublet is
necessary. Even from the partial calculation\cite{nmsgut3} we see
that the large value of the wave function dressing makes the GUT
tree level matter fermion yukawa couplings (i.e
$\{h_{AB},g_{AB},f_{AB}\}_{tree}$ and therefore all the
$(y^f_{AB})_{tree}$ ) required to match the SIMSSM couplings at
$M_X^0$ much  smaller than they would be without these
corrections.  This has the important consequence of suppressing
$d=5$   B-violation operators since they depend on these yukawas
and have no Higgs line.  Since it is the tree level couplings that
enter the formulae for the inflaton dynamics in the full GUT it is
easier to satisfy eqn.(\ref{hsqbyM}). Because of this and the
relatively large value of  $M\sim M_3$ it should be be possible to
achieve the required fine tuning using  the full wave function
dressing\cite{csaigck}.

Embedding in the GUT has overturned our naive assumption  that the
lowest intermediate scale would govern inflation. Instead it is
rather the largest. While setting us the  problem of finding
solutions to the tuning condition, compatible both  with an
accurate fit of fermion masses and acceptable values of
inflationary power spectrum and spectral index, it emphatically
shows that the soft terms have little role to play in the fine
tuning which belongs rather to the GUT and intermediate scale
physics only. Thus the physics of SIMSSM driven inflation is in
sharp contrast to the Dirac neutrino mass driven
inflation\cite{akm,hotchmaz} in the MSSM extended by $U(1)_{B-L}$
and right handed neutrinos. Our analysis makes it clear that they
lie counterpoised  not only as regards the nature of neutrino mass
but also as regards the nature of inflation and its regulating
mass scale besides their degree of naturalness.

 An example of the relevant parameters from  an accurate fit  of
the complete fermion spectrum in the NMSGUT  which has also been
tuned to make it as compatible as possible with the inflationary
scenario presented is seen in Table 1. More details  may be found
in \cite{ssi}. The fine tuning between the yukawas proceeds as
anticipated with $1-\Gamma =1=\Lambda_{n,d}$. The remaining
problem is  that $h^2/M\sim 10^{-19}$ GeV is too large. As a
result the number of e-folds $N_{CMB}$ is much smaller than
required. However as explained the formulae used seriously
underestimate the Higgs wave function corrections.
  Search of the huge parameter space has just begun. We may well
  hope for  a  completely realistic fit compatible with inflation in due
  course. The detailed analysis of reheating behaviour and
  Lepto-genesis in this model is also underway. Since the fits of
   fermion masses  also yield a value for the CP violation parameter
  relevant for Lepto-genesis the incorporation of the SSI scenario
   in the NMSGUT may eventually yield useful additional
   constraints which will serve to narrow the parameter space further.

 \begin{table}
 $ \begin{array}   {|c|c|c|c|}
\hline
 {\rm  Parameter }&Value &Parameter  &Value\\
 \hline
 \chi_{X}&        0.4458   & M_{h^0}&  122.99 \\
 \chi_{Z}&         0.1426 &   M_X & 7.08 \times 10^{17}\\
     f_3  &  1.066 \times {10}^{-3}& f_1 ,f_2 &2.59\times{10}^{-8},4.405 \times 10^{-5}   \\
     h & 2.44 \times 10^{-4} & \Lambda_{n} &0.999999\\
     M & 3.043 \times 10^{11}& \Lambda_{d} & 0.999999\\
     \Gamma & 4.343 \times 10^{-5}& \Delta_{tuning}& 0.989\\
     |{\overline{\sigma}}| & 4.69 \times 10^{15} &M_X &5.25 \times 10^{17}\\
A_0(M_X),m_0(M_X) & -5.235\times 10^5 , 1.260 \times 10^4 &\mu,B(M_X)  & 4.316\times 10^5,-1.128 \times 10^{11}\\
M^2_{\bar H} &   -1.498 \times 10^{11} & M^2_H & -1.448 \times 10^{11}\\
|\Delta_{H_0}|,|\Delta_{\bar{H}_0}| & 50.254,63.930 & |\alpha_{4}| & 0.707\\
M^{\nu^c}_3 & 4.86  \times 10^{13}&M^{\nu^c}_{1,2}  & 1.181 \times 10^{9},2.01\times 10^{12}  \\
 |y_{31tree}^\nu|& 1.997 \times 10^{-4}& |y_{21tree}^\nu|,|y_{11tree}^\nu| & 4.489 \times 10^{-5},1.640 \times 10^{-6}\\
 Log_{10}(h^2/M) & -18.706 & V_0,\phi_{end} & 3.579 \times 10^{52},2.153 \times 10^{15}\\
 N_{pivot},N_{CMB} & 54.22, 4.78 \times 10^{-4}  & \Delta , \beta & 8.82 \times 10^{-12},5.92 \times10^{-6}  \\
      \hline
 \end{array}
 $
 \label{table I}\caption{\small{Illustrative example  of relevant parameters from an accurate fit
 of the fermion spectrum in the NMSGUT which is compatible with inflationary
  scenario. All masses are in GeV.
  $\chi_{X,Z}$ are the accuracies of the fits to 18 known fermion mass/mixing  parameters
   at $M_{X,Z}$.}}
 \end{table}


\begin{theacknowledgments}
 This is a report of work done in collaboration with Ila Garg and published as \cite{ssi}.
   We are grateful to Anupam Mazumdar and  Ling Fei Wang
 for correspondence and collaboration in earlier stages of this  work.
and  thank  David Lyth for discussions and useful comments.

\end{theacknowledgments}



 \bibliographystyle{aipproc}   




\end{document}